\documentclass[acmsmall,screen]{acmart}

\setcopyright{acmlicensed}
\copyrightyear{2026}
\acmYear{2026}
\acmConference[CP'26]{2nd International Workshop on Choreographic Programming}{June 15--19,
  2026}{Boulder, CO}

\usepackage{xspace}
\usepackage{comment}
\usepackage[frozencache]{minted}
\usepackage{booktabs}
\usepackage{pifont}
\usepackage{multirow}
\usepackage{wrapfig}
\usepackage{tikz}
\usetikzlibrary{
  positioning, fit, calc, arrows.meta,
  backgrounds, shapes.geometric, shapes.misc,
  tikzmark
}
\usepackage{pgfplots}
\pgfplotsset{compat=1.18}
\usepackage{enumitem}
\setlist[itemize]{leftmargin=1.75em}

\definecolor{shadecolor}{gray}{1.00}
\definecolor{darkgray}{gray}{0.30}
\definecolor{lightgray}{gray}{0.91}
\definecolor{mygray}{rgb}{0.5,0.5,0.5}

\definecolor[named]{ACMBlue}{cmyk}{1,0.1,0,0.1}
\definecolor[named]{ACMDarkBlue}{cmyk}{1,0.58,0,0.21}
\definecolor[named]{ACMOrange}{cmyk}{0,0.42,1,0.01}
\definecolor[named]{ACMRed}{cmyk}{0,0.90,0.86,0}
\definecolor[named]{ACMGreen}{cmyk}{0.20,0,1,0.19}
\definecolor[named]{ACMPurple}{cmyk}{0.55,1,0,0.15}

\newcommand{\ie}{\emph{i.e.}\xspace}

\newcommand{\cf}{\textit{cf.}\xspace}

\newcommand{\tname}[1]{\textsc{#1}\xspace}
\newcommand{\langname}{\tname{Pact}}

  {\unskip\nobreak\hskip 1em plus 1fil\nobreak$\square$
   \parfillskip=0pt%
   \endtrivlist}

\newcommand{\hide}[1]{}

\definecolor{basisBlue}{RGB}{56,104,255}
\definecolor{basisBlueLite}{RGB}{230,238,255}
\definecolor{basisGreen}{RGB}{60,170,95}
\definecolor{basisGreenLite}{RGB}{226,247,233}
\definecolor{basisGray}{RGB}{120,130,150}
\definecolor{basisGrayLite}{RGB}{245,247,252}
\definecolor{panelBlue}{RGB}{120,170,230}
\definecolor{scriptOrange}{RGB}{224,148,60}
\definecolor{propPurple}{RGB}{145,95,255}
\definecolor{propPurpleLite}{RGB}{244,238,255}

\tikzset{
  highlightbox/.style={dashed, rounded corners},
  old/.style={highlightbox, fill=ACMRed!5, draw=ACMRed, inner sep=0.0em},
  new/.style={highlightbox, fill=ACMGreen!5, draw=ACMGreen},
  moveline/.style={->, line width=.5mm, dashed, >=stealth},
  changeline/.style={moveline, draw=ACMGreen}
}
\tikzset{
  labelbox/.style={draw=black!60, rounded corners=1pt,
                   minimum width=14mm, minimum height=6mm,
                   font=\footnotesize, inner sep=2pt}
}

\tikzset{
  headerGray/.style={draw=basisGray!55, fill=basisGrayLite, rounded corners=3pt,
                     minimum width=20mm, minimum height=7mm, line width=0.6pt, inner xsep=0pt},
  headerBlue/.style={draw=basisBlue, fill=basisBlueLite, rounded corners=3pt,
                     minimum width=20mm, minimum height=7mm, line width=0.8pt},
  headerGreen/.style={draw=basisGreen, fill=basisGreenLite, rounded corners=3pt,
                      minimum width=20mm, minimum height=7mm, line width=0.8pt},
  lifelineGray/.style={draw=basisGray!35, dashed, line width=0.8pt},
  lifelineBlue/.style={draw=basisBlue!35, dashed, line width=0.9pt},
  lifelineGreen/.style={draw=basisGreen!35, dashed, line width=0.9pt},
  msgBlue/.style={draw=basisBlue, line width=0.9pt, -{Stealth[length=2.2mm,width=2.2mm]}},
  msgGreen/.style={draw=basisGreen, line width=0.9pt, -{Stealth[length=2.2mm,width=2.2mm]}},
  msgBlack/.style={draw=basisGray!95, line width=0.9pt, -{Stealth[length=2.2mm,width=2.2mm]}},
  evalbox/.style={draw=scriptOrange, dashed, rounded corners=2pt,
                  line width=0.9pt, inner xsep=6pt, inner ysep=4pt,
                  text=scriptOrange!90!black},
  propbox/.style={draw=propPurple, fill=propPurpleLite, rounded corners=3pt,
                  line width=0.9pt, inner xsep=5pt, inner ysep=5pt,
                  text=propPurple!90!black},
  phaseSeg/.style={rounded corners=2pt},
  phaseLabel/.style={rotate=90, anchor=center, font=\bfseries\scriptsize}
}

\hypersetup{
  colorlinks,
  linkcolor=ACMDarkBlue,
  citecolor=ACMPurple,
  urlcolor=ACMDarkBlue,
  filecolor=ACMDarkBlue
}

\definecolor{agentfill}{HTML}{2D9B9B}
\definecolor{panelborder}{HTML}{4A90D9}
\definecolor{gadgetfill}{HTML}{B8CCE4}
\definecolor{orangeborder}{HTML}{E0A030}
\definecolor{orangefill}{HTML}{F0B84D}
\definecolor{docfill}{HTML}{F5E6C8}
\definecolor{bluebox}{HTML}{4A90D9}
\definecolor{lockbody}{HTML}{D4A030}
\definecolor{lockshackle}{HTML}{B8902A}

\tikzset{
  agent/.style={
    regular polygon, regular polygon sides=3,
    draw=black, fill=agentfill, line width=0.5mm,
    minimum size=10mm, inner sep=0pt,
  },
  smalldoc/.style={
    rectangle, minimum width=8mm, minimum height=2.8mm, inner sep=0pt,
  },
  bluedoc/.style={smalldoc, draw=bluebox!60, fill=bluebox!20, line width=0.5mm},
  yellowdoc/.style={smalldoc, draw=orangeborder!60, fill=docfill, line width=0.5mm},
  gadget/.style={
    rectangle, draw=bluebox, fill=gadgetfill,
    minimum width=42mm, minimum height=14mm,
    font=\bfseries\small, rounded corners=1pt, line width=0.5mm
  },
  outbox/.style={
    rectangle, draw=bluebox, fill=white,
    minimum width=34mm, minimum height=8.5mm,
    font=\small, rounded corners=1pt, line width=0.5mm
  },
  proofbox/.style={
    rectangle, draw=bluebox, fill=gadgetfill!40,
    minimum width=34mm, minimum height=8.5mm,
    font=\small, rounded corners=1pt, line width=0.5mm
  },
  scriptbox/.style={
    rectangle, draw=orangeborder, fill=white,
    minimum width=21mm, minimum height=7.5mm,
    font=\small, rounded corners=1pt, line width=0.5mm
  },
  zk/.style={
    circle, draw=black, fill=orangefill,
    minimum size=10mm, font=\footnotesize\bfseries, line width=0.5mm
  },
  panel/.style={
    draw=panelborder, dashed, line width=1.2pt,
    rounded corners=6pt, inner sep=0pt
  },
  paneltitle/.style={font=\Large\bfseries},
  arr/.style={-{Stealth[length=2.5mm]}, black, line width=1.5pt, >=stealth},
  net/.style={black, line width=1.5pt}
}

\begin{document}

\title{Pact: A Choreographic Language for Agentic Ecosystems}

\author{Kiran Gopinathan}
\email{kiran@basis.ai}
\orcid{https://orcid.org/0000-0002-1877-9871}
\affiliation{%
 \institution{Basis Research Institute}
 \city{New York}
 \state{New York}
 \country{USA}
}
\author{Jack Feser}
\email{jack@basis.ai}
\orcid{https://orcid.org/0000-0001-8577-1784}
\affiliation{%
 \institution{Basis Research Institute}
 \city{New York}
 \state{New York}
 \country{USA}
}
\author{Michelangelo Naim}
\email{michi@basis.ai}
\orcid{https://orcid.org/0000-0002-4419-9301}
\affiliation{%
 \institution{Basis Research Institute}
 \city{New York}
 \state{New York}
 \country{USA}
}
\author{Zenna Tavares}
\email{zenna@basis.ai}
\affiliation{%
 \institution{Basis Research Institute}
 \city{New York}
 \state{New York}
 \country{USA}
}
\author{Eli Bingham}
\email{eli@basis.ai}
\affiliation{%
 \institution{Basis Research Institute}
 \city{New York}
 \state{New York}
 \country{USA}
}



\renewcommand{\shortauthors}{Gopinathan et al.}

\begin{abstract}
  Recent advances in large language models have led to the rise of
  software systems (\ie~agents) that execute with increasing autonomy
  on behalf of users in open, multi-party settings, interacting with
  untrusted counterparts and managing private information.
  Choreographic programming offers correct-by-construction
  protocol-design for such settings, but assumes cooperative
  participants --- it has no notion of agent self-interest, that is,
  \emph{why} an agent will follow a protocol.
  In this talk we introduce \langname, a choreographic language
  extended with operations to describe agent choices and preferences,
  drawing from the rich literature of game theory.
  Every \langname protocol maps to a formal game, allowing protocol
  designers to reason about game-theoretic properties of their
  protocols, such as solving for decision policies.
  We present \langname's design and a preliminary implementation --- a
  bounded-rational solver that computes decision policies over
  \langname protocols --- and findings from applying this language to
  multi-party coordination with self-interested agentic participants.

\end{abstract}




\received{30 March 2026}

\maketitle

\section{Introduction}
Advances in large language models have led to the rise of autonomous
agents that execute increasingly autonomously on behalf of users in
open, multi-party settings, interacting with untrusted counterparts
and managing private information.
While LLM-based AI is capable, in practice it is rarely trustworthy,
especially when it must negotiate with unpredictable, possibly
hostile, other agents.
Evidence of this is already abundant: agents are systematically
vulnerable to prompt injection
attacks~\cite{liuetalpromptinjection,agentdojoneurips24}, they
regularly exhibit sycophantic behaviour that undermines rational
reasoning, and they can easily be coerced into acting against their
own best interests~\cite{AgentsRuleTwo,MultiagentAISystems2026}.
If these ecosystems are to be sustainable, agents urgently need tools
to manage these interactions.

Trust in multi-party systems fundamentally boils down to coordination,
and problems of coordination are well studied in the PL literature.
Choreographic programming~\cite{montesiIntroductionChoreographies2023}
describes one such formalism for this purpose, designed in such a way
that choreographic programs are correct by construction---guaranteed
to be \emph{deadlock-free}.
Implementations and embeddings of choreographic programs now exist in
several production languages
~\cite{shenHasChorFunctionalChoreographic2023,giallorenzoChoralObjectorientedChoreographic2024,
  batesEfficientPortableCensusPolymorphic2025} and the area is rapidly
maturing.
Recent work has begun investigating enriching them with various
cryptographic primitives~\cite{sarkarHasTEEConfidentialCloud2024,
  veigaChoreographiesSecureComputation2025}, extending them towards
adversarial settings where participants may not be trusted to execute
faithfully, opening the door to the use of choreographies to manage
multi-agentic coordination.

\begin{figure}[t]
\begin{minipage}[t]{0.30\columnwidth}
\centering\textbf{Choreography}
\begin{minted}[fontsize=\scriptsize]{ocaml}
(* title  : string @ buyer,
   budget : int @ buyer
   book   : Book @ seller *)
let bookseller () =
  send(title, seller)
  price = seller.price_of(title)
  send(price, buyer)
  if broadcast(price < budget)
  then exchange(
      buyer, seller,
      book, price
   )
\end{minted}
\end{minipage}%
\hfill\vrule\hfill%
\begin{minipage}[t]{0.30\columnwidth}
\centering\textbf{Buyer}
\begin{minted}[fontsize=\scriptsize]{ocaml}
(* title  : string,
   budget : int     *)

let bookseller_buyer () =
  send(title, seller)
  price = recv(seller)
  choice = price < budget
  send(choice, seller)
  if choice then begin
    book = recv(seller)
    book = send(price, seller)
  end
\end{minted}
\end{minipage}%
\hfill\vrule\hfill%
\begin{minipage}[t]{0.30\columnwidth}
\centering\textbf{Seller}
\begin{minted}[fontsize=\scriptsize]{ocaml}
(* book : Book *)


let bookseller_seller () =
  title = recv(buyer)
  price = price_of(title)
  send(price, buyer)
  choice = recv(buyer)
  if choice then begin
    send(book, buyer)
    balance += recv(buyer)
  end
\end{minted}
\end{minipage}
\caption{A choreographic bookseller protocol (left) and its endpoint
  projections for buyer (middle) and seller (right). The choreography
  compiles to local programs, guaranteeing deadlock-freedom by
  construction.}
\label{lst:bookseller}
\end{figure}

Could choreographies be used to reason about agentic interactions?
Consider Figure~\ref{lst:bookseller} (left), which describes a
canonical choreography ``hello world''---a book seller protocol.
The choreography describes an interaction between two parties---a
buyer and a seller: the buyer sends a title to the seller, and the
seller, computing a price locally, responds back to the buyer. If the
price is within budget, both parties conduct an exchange and the book
is sold.
The choreographic language guarantees deadlock freedom, and this
choreography can be compiled via endpoint projection to local
programs where every send is matched by a corresponding receive by the
other parties.

While this choreography captures the communication pattern between the
two parties precisely, this program fails to capture a crucial aspect
of agentic interactions: \emph{why} would any agent follow this
protocol?
Agents operating in such ecosystems have conflicting objectives, \ie a
buyer wants the cheapest price for a good, the seller wants the most
profit, and will only participate in protocols that serve their objectives.
Choreographies specify the structure of interaction, but have no model
of preferences, and so provide no help in reasoning about such questions.

In this paper, we present \langname, a choreographic language that
adds three constructs to facilitate agentic
coordination: \emph{explicit agent choices}, where agents make
strategic decisions within a protocol (\ie, what price to sell a book
at, what price to purchase); \emph{explicit agent utilities}, where
agents declare what they value about outcomes (\ie, how much an agent
values the quality of the book they receive), and \emph{nature
  variables}, which model agent priors about the world (\ie
assumptions about the distribution of book quality).
Every \langname protocol maps unambiguously to a formal game, and
agent preferences and decisions are made explicit and targets of
formal reasoning.
Agents can use \langname protocols as a medium for negotiation,
proposing and analysing these programs collaboratively to facilitate
large scale ecosystems of autonomous agents.

In summary, the key contributions of this work are:
\begin{itemize}
\item Identification of game-theoretic reasoning as a natural extension
  of choreographic programming, enabling its application to the rapidly emerging domain of multi-agentic coordination.
\item The initial design of \langname, a choreographic language with semantics grounded in game theory.
\item An example game-theoretic decision policy solver built as an analysis on top of \langname programs.
\end{itemize}

The rest of this paper is organised as follows:
Section~\ref{sec:taste-pact} walks through the design of \langname by
example, extending the bookseller choreography with choices, utilities
and nature variables;
Section~\ref{sec:memo} presents a game-theoretic
analysis built on top of \langname, using theory-of-mind models to
construct a bounded-rational decision policy solver for \langname programs;
Section~\ref{sec:related} presents a brief survey of the relevant
literature around formalising games and agentic coordination, and
Section~\ref{sec:conclusion} concludes this work and outlines
directions of future work we are investigating.


\section{How to Forge a \langname}
\label{sec:taste-pact}
We now walk through the design of \langname by extending the
bookseller choreography (\cf~Figure~\ref{lst:bookseller})
incrementally to address concerns relevant in multi-agentic coordination.
Each extension extends the language to capture an additional aspect of
agentic interaction: (1) explicit modeling of agent choices, (2)
encoding of agent utilities, and (3) representation of priors using
nature variables.
By the end of this section, the bookseller protocol in \langname will
correspond to a complete formal game.

\subsection{Making Strategic Choices Explicit}\label{sec:taste-choices}

\begin{wrapfigure}{r}{0.33\columnwidth}
\vspace{-1.2em}
\begin{tikzpicture}[remember picture, overlay]
  \draw[rounded corners=2pt, fill=ACMBlue!12, draw=none, thick]
    ([shift={(-2pt,1.6ex)}]pic cs:choice1s)
    rectangle
    ([shift={(2pt,-0.4ex)}]pic cs:choice1e);
  \draw[rounded corners=2pt, fill=ACMBlue!12, draw=none, thick]
    ([shift={(-2pt,1.6ex)}]pic cs:choice2s)
    rectangle
    ([shift={(2pt,-0.4ex)}]pic cs:choice2e);
\end{tikzpicture}
\begin{minted}[fontsize=\scriptsize,escapeinside=@@]{ocaml}
let bookseller () =
  send(title, seller)
  price = @\tikzmark{choice1s}@seller.choose(float)@\tikzmark{choice1e}@
  send(price, buyer)
  if broadcast(@\tikzmark{choice2s}@buyer.choose(bool)@\tikzmark{choice2e}@)
  then exchange(
      buyer, seller,
      book, price
   )
\end{minted}
\vspace{-1em}
\end{wrapfigure}
To make our choreographies useful for agentic interactions, we must be
able to analyse them as games,
and the core of any game is about the strategic choices of its
participants --- how each participant will make choices taking into
consideration the actions of others.
Our initial book seller choreography in some sense is \emph{too}
prescriptive: it states exactly how each agent will compute its local
choices, that the seller will use the price given by its function
\mintinline{python}{price_of}, and the buyer will purchase the good if
the price is less than its budget.
But is this accurate in an agentic setting? Can we trust the seller to
not adjust the price based on their beliefs about the buyer?
Similarly, will the buyer purchase the book for \emph{any} price
within budget?
To capture the strategic nature of this interaction, the first thing
we must do is make the \emph{choices} of each participant explicit: we
replace each agent's hardcoded choices with a
\mintinline{python}{agent.choose} method. In terms of the dynamic
semantics of the language, we model these as non-deterministic
choices, underspecified by the language, but relevant for
game-theoretic analyses.

\subsection{Encoding Agent Utilities}\label{sec:taste-values}
\begin{wrapfigure}{r}{0.32\columnwidth}
\vspace{-1.2em}
\begin{tikzpicture}[remember picture, overlay]
  \draw[rounded corners=2pt, fill=ACMBlue!15, draw=none, thick]
    ([shift={(-2pt,1.6ex)}]pic cs:values1s)
    rectangle
    ([shift={(2pt,-0.4ex)}]pic cs:values1e);
\end{tikzpicture}
\begin{minted}[fontsize=\scriptsize,escapeinside=@@]{ocaml}
let bookseller () =
  @\tikzmark{values1s}@buyer.values(book.quality)@\tikzmark{values1e}@
  send(title, seller)
  price = seller.choose(float)
  send(price, buyer)
  if broadcast(buyer.choose(bool))
  then exchange(
      buyer, seller,
      book, price
   )
\end{minted}
\vspace{-1em}
\end{wrapfigure}
Our choice operator tells us \emph{what} agents decide, but not
\emph{why}.  
In our original protocol, the value gained by the seller through
participating in the protocol was clear, the
\mintinline{python}{price} it received from the buyer as the fourth
argument to \mintinline{python}{exchange}, but what does the
\emph{buyer} gain?
To reason about such strategic decisions, our language needs some
model of agent objectives.
We draw from game theory and equip each agent with utility functions
over outcomes, encoding our beliefs of what each participant stands to
gain or lose from a protocol's execution.
We assume an abstract notion of money which is valued by all
participants, capturing the monetary gain of the seller.
The \mintinline{python}{values} operator declares an agent's utility:
writing \mintinline{python}{buyer.values(book.quality)} states that
the buyer's payoff from this interaction also depends on the quality
of the book.
Notably, the \mintinline{python}{values} operator only states factors
influencing the utility function, not the nature of this relation ---
agents' true utility functions are of strict strategic relevance, and
no reasonable agent could be persuaded to report them honestly.
Instead the declaration serves as a basis for negotiation, that is, by
stating what each participant values, agents can propose or
counter-propose protocols that others are incentivised to accept. A
buyer that fails to declare value on the book invites protocols where
it pays and receives nothing.
The \mintinline{python}{values} declarations establish terms on which
protocols are mutually acceptable.
Once added, each agent can inject beliefs about other participants'
utilities into the protocol and analyse whether participating is
beneficial.

\subsection{Capturing Priors over the World}\label{sec:taste-nature}

So far, we have extended \langname by making agent decisions and
objectives explicit, but one more component is needed to capture the
full strategic nature of agentic interactions: prior assumptions about
the world also substantially influence how agents reason and act.
A buyer that believes books are generally high quality might accept a
higher price, while one who does not, will refuse.
To capture these additional factors, we extend every protocol with a
third participant, \emph{the world}, whose decisions correspond to
exogenous actions, that is, those that must be modelled independently
by each participant.
In particular, in the book seller protocol, we update the protocol to
declare that the quality of the book is a property determined outside
of the actions of each participant, \texttt{book.quality <-
  world.choose(float)}.
Similar to the value declaration, these serve as a basis for
negotiation, and when analysing the game, agents will inject their
priors over the world and use them to reason about the expected payoff
from participating.

\subsection{Pact, a language for protocols with Trust}
\label{sec:pact-lang-prot}
\begin{wrapfigure}[13]{r}{0.40\columnwidth}
\begin{tikzpicture}[remember picture, overlay]
  \draw[rounded corners=2pt, fill=ACMBlue!18, draw=none, thick]
    ([shift={(-2pt,1.6ex)}]pic cs:nature1s)
    rectangle
    ([shift={(2pt,-0.4ex)}]pic cs:nature1e);
\end{tikzpicture}
\begin{minted}[fontsize=\scriptsize,escapeinside=@@]{ocaml}
let bookseller () =
  @\tikzmark{nature1s}@book.quality <- world.choose(float)@\tikzmark{nature1e}@
  buyer.values(book.quality)
  send(title, seller)
  price = seller.choose(float)
  send(price, buyer)
  if broadcast(buyer.choose(bool))
  then exchange(
      buyer, seller,
      book, price
   )
\end{minted}
\caption{\langname encoding of the full buyer seller protocol, with explicit agent choices, preferences and priors over the world.}
\label{fig:pact-buyer-seller}
\end{wrapfigure}
Putting it all together Figure~\ref{fig:pact-buyer-seller} presents
the canonical buyer seller protocol encoded as a \langname protocol.
The protocol first declares that the quality of the book will be
modelled as an exogenous variable, and that the buyers utility will be
dependent on this quality.
The protocol then proceeds as a traditional choreography, with the
buyer sending the title it wants to the seller.
We replace local computations with strategic relevance with explicit
choice operators, so in the next step, the seller chooses a price and
sends it back to the buyer --- each strategic operation is assumed to
be made taking into account any and all information that the agent has
access to, \ie in this case, the title of the book.
The buyer then makes a decision about this price, and broadcasts it to
all participants to branch.
If the price is accepted, then the exchange concludes and the buyer
receives the book, the seller the price.
In this way, \langname protocols capture both the communication
pattern of a distributed protocol, enough strategic information to
encode a formal game.
This allows participants to analyse these protocols to answer game
theoretic questions of their interactions, such as whether
participating in a protocol is beneficial given their assumptions, or
as we will demonstrate in the next section, decision policies for
their local choices.


\section{A decision policy solver for \langname protocols}
\label{sec:memo}
\langname protocols correspond to formal games, and can be analysed as such.
In this section we present an initial such analysis built on top of
\langname, implementing a decision policy solver that computes how
agents should act given their beliefs about each other. 
In doing so, we obtain a rather fun result: our canonical
choreographic bookseller protocol, treated as a game, allows us to
recreate the game theory phenomenon of a Market for
Lemons~\cite{Akerlof70}, where buyers, uncertain of the quality of the
good, must reason through information asymmetry with only access to
the price.

\subsection{Theory of Mind for Protocols}
How should agents make their local decisions in a \langname protocol?
What price \emph{should} the seller set for its books? 
What price should the buyer make a purchase?
The answer to these questions is in some sense inherently recursive.
The price the seller should set is the highest price the buyer will
accept; the price at which the buyer should accept is the lowest price
the seller will set for a high quality good, and so on.
Given priors on exogenous variables and utility functions, solving
for a decision policy requires modelling other agents' reasoning, that
is, a \emph{theory of mind}.

\begin{figure}
\centering
\definecolor{tomBlue}{HTML}{6670B8}      
\definecolor{tomLavender}{HTML}{B8A0D8}  
\definecolor{tomSellerBg}{HTML}{D6DAF0}  
\definecolor{tomSellerBd}{HTML}{9AA0D0}  
\definecolor{tomBuyerBg}{HTML}{E8D8F0}   
\definecolor{tomBuyerBd}{HTML}{BCA8D0}   
\definecolor{tomWorldBg}{HTML}{F0EDE5}   
\definecolor{tomWorldBd}{HTML}{C8C0B0}   
\begin{tikzpicture}[
    known/.style={circle, fill=tomBlue, text=white,
                  minimum size=13pt, font=\tiny\bfseries,
                  inner sep=0pt},
    uncertain/.style={circle, fill=tomLavender, text=white,
                      minimum size=13pt, font=\tiny\bfseries,
                      inner sep=0pt},
    util/.style={diamond, draw=black!35, fill=black!4,
                 minimum size=10pt, font=\tiny\bfseries,
                 inner sep=1pt, line width=0.4pt},
    sellerframe/.style={draw=tomSellerBd, fill=none,
                        rounded corners=3pt, line width=0.5pt,
                        minimum width=2.6cm, minimum height=1.1cm},
    sellerframe-sim/.style={sellerframe, dashed},
    buyerframe/.style={draw=tomBuyerBd, fill=none,
                       rounded corners=3pt, line width=0.5pt,
                       minimum width=2.6cm, minimum height=1.1cm},
    buyerframe-sim/.style={buyerframe, dashed},
    worldframe/.style={draw=black!50, fill=none,
                       rounded corners=3pt, line width=0.5pt,
                       minimum width=2.6cm, minimum height=1.1cm},
    dep/.style={->, >=stealth, black!60, line width=0.4pt},
    cond/.style={->, >=stealth, black!50, densely dashed, line width=0.4pt},
    recur/.style={->, >=stealth, black!50, densely dashed, line width=0.5pt, draw=tomLavender!10!tomBlue},
    lbl/.style={font=\tiny, text=black, fill=white!20, rounded corners},
    sellerlbl/.style={font=\tiny\bfseries, text=black},
    buyerlbl/.style={font=\tiny\bfseries, text=black},
    wframelbl/.style={font=\tiny\bfseries, text=black},
    title/.style={font=\footnotesize\bfseries},
  ]

  \def\yrowA{2.4}
  \def\yrowB{0.7}
  \def\yworld{3.8}

  \def\xleft{-2.6}
  \def\xright{2.8}
  \def\nodeoff{0.8}

  \node[title] at (\xleft, 4.6) {Buyer model};

  \node[uncertain] (bwq) at (\xleft, \yworld) {$q$};

  \node[known] (bsq) at ({\xleft-\nodeoff}, \yrowA) {$q$};
  \node[lbl, above=0.5pt of bsq] {knows};
  \node[uncertain] (bsp) at ({\xleft+\nodeoff}, \yrowA) {$p$};
  \node[lbl, above=0.5pt of bsp] {chooses};
  \draw[dep] (bsq) -- (bsp);

  \draw[dep] (bwq) -- (bsq);

  \node[known] (bbp) at ({\xleft-\nodeoff}, \yrowB) {$p$};
  \node[lbl, above=1pt of bbp] {observes};
  \node[uncertain] (bbq) at ({\xleft+\nodeoff}, \yrowB) {$\hat{q}$};
  \node[lbl, above=1pt of bbq] {infers};
  \draw[dep] (bbp) -- (bbq);

  \draw[cond, out=-100, in=80] (bsp) to (bbp);

  \node[util] (bu) at (\xleft, -0.3) {$U$};
  \node[lbl, below=1pt of bu, text=black] {$\hat{q} - p$};
  \draw[dep] (bbq) -- (bu);
  \draw[dep] (bbp) -- (bu);

  \node[title] at (\xright, 4.6) {Seller model};

  \node[known] (ssq) at ({\xright-\nodeoff}, \yrowA) {$q$};
  \node[lbl, above=1pt of ssq] {knows};
  \node[uncertain] (ssp) at ({\xright+\nodeoff}, \yrowA) {$p$};
  \node[lbl, above=1pt of ssp] {chooses};
  \draw[dep] (ssq) -- (ssp);

  \node[known] (sbp) at ({\xright-\nodeoff}, \yrowB) {$p$};
  \node[lbl, above=1pt of sbp] {observes};
  \node[uncertain] (sbq) at ({\xright+\nodeoff}, \yrowB) {$\hat{q}$};
  \node[lbl, above=1pt of sbq] {infers};
  \draw[dep] (sbp) -- (sbq);

  \draw[cond, out=-100, in=80] (ssp) to (sbp);

  \node[util] (su) at (\xright, -0.3) {$U$};
  \node[lbl, below=1pt of su, text=black] {$p \cdot \Pr[\hat{q}{=}1]$};
  \draw[dep] (ssp) -- ++(0.9, 0) |- (su);
  \draw[dep] (sbq) -- (su);

  \begin{scope}[on background layer]
    \node[worldframe, fit=(bwq),
          inner xsep=8pt, inner ysep=6pt,
          label={[wframelbl,anchor=east]left:world}] (bw) {};
    \node[sellerframe-sim, fit=(bsq)(bsp),
          inner xsep=8pt, inner ysep=6pt,
          label={[sellerlbl,anchor=south east]north west:seller}] (bs) {};
    \node[buyerframe, fit=(bbp)(bbq),
          inner xsep=8pt, inner ysep=6pt,
          label={[buyerlbl,anchor=south east]north west:buyer}] (bb) {};
    \node[sellerframe, fit=(ssq)(ssp),
          inner xsep=8pt, inner ysep=6pt,
          label={[sellerlbl,anchor=south east]north west:seller}] (ss) {};
    \node[buyerframe-sim, fit=(sbp)(sbq),
          inner xsep=8pt, inner ysep=6pt,
          label={[buyerlbl,anchor=south east]north west:buyer}] (sb) {};
  \end{scope}

  \path (bs.east) -- (ss.west) coordinate[midway] (recmidA);
  \draw[recur, rounded corners=3pt]
    (bs.east) -- (ss.west);
  \node[lbl, fill=white, inner sep=1pt] at (recmidA) {$\ell{-}1$};

  \path (sb.west) -- (bb.east) coordinate[midway] (recmidB);
  \draw[recur, rounded corners=3pt]
    (sb.west) --  (bb.east);
  \node[lbl, fill=white, inner sep=1pt] at (recmidB) {$\ell{-}1$};

  \node[known, minimum size=7pt] at (-3.6, -1.3) {};
  \node[lbl, anchor=west] at (-3.3, -1.3) {known};
  \node[uncertain, minimum size=7pt] at (-1.9, -1.3) {};
  \node[lbl, anchor=west] at (-1.6, -1.3) {uncertain};
  \node[util, minimum size=6pt, inner sep=0pt] at (-0.1, -1.3) {};
  \node[lbl, anchor=west] at (0.15, -1.3) {utility};
  \draw[cond] (1.2, -1.3) -- ++(0.4, 0);
  \node[lbl, anchor=west] at (1.7, -1.3) {cond.};
  \draw[recur] (2.6, -1.3) -- ++(0.4, 0);
  \node[lbl, anchor=west] at (3.1, -1.3) {simulates};

\end{tikzpicture}
\caption{Recursive theory-of-mind structure for the buyer-seller
  protocol. Each agent maintains a generative model of the interaction:
  \textbf{Left:} the buyer imagines a world that draws book quality,
  and a seller (dashed frame) who observes it and sets a price; the
  buyer then conditions on the observed price to infer quality and
  evaluates utility as inferred quality minus price.
  \textbf{Right:} the seller knows quality, chooses a price, and
  simulates a buyer (dashed frame) who observes that price and infers
  quality; the seller maximises expected revenue weighted by the
  probability the simulated buyer believes quality is high.
  Dashed arrows between frames mark recursive calls at depth
  $\ell{-}1$, bottoming out at $\ell = 0$ with na\"ive priors.
  Solid frames denote the modelling agent's own perspective; dashed
  frames denote simulated counterparts.}
\label{fig:tom-recursion}
\end{figure}
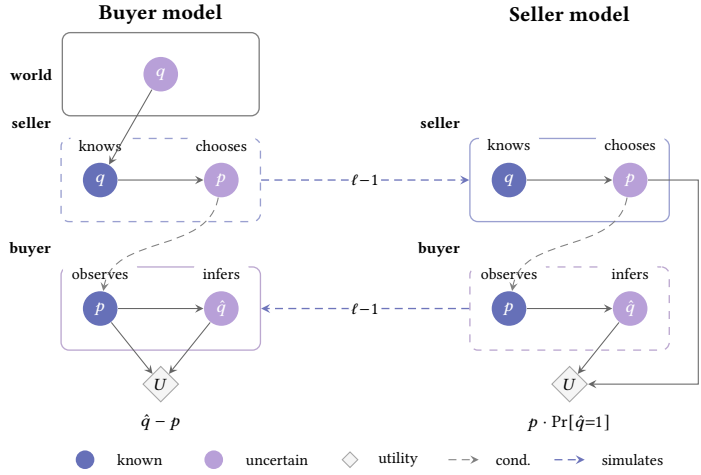

We implement an analysis based on this observation using the
\texttt{memo} programming
language~\cite{chandraReasoningReasoning2025}.
The \texttt{memo} language was developed to aid cognitive science
researchers build efficient computational models of theory of mind,
providing a concise DSL for expressing such models, and a compiler
that reduces them to recursive probabilistic programmes over which
inference can be performed efficiently.
Figure~\ref{fig:tom-recursion} illustrates such a recursive structure 
that could be used to model the bookseller protocol.
In this model, the buyer (left) makes decisions with a theory of mind
by simulating a world and a seller, and conditions on the observed
price to make an inference about quality;
the seller (right) similarly simulates a buyer and chooses a price
that maximises expected revenue.
Each party in the interaction recursively simulates the
other. Depending on the depth of this analysis, these recursive models
themselves contain nested, \emph{even smaller} recursive models.

Our key insight here is that a \langname protocol contains all the
information to precisely construct theory-of-mind models, and we use
this to build a decision policy solver for \langname games.
Figure~\ref{fig:memo-buyer} illustrates a theory-of-mind program
constructed purely through an endpoint projection of the buyer-seller
pact protocol (\cf~\ref{fig:pact-buyer-seller}).
The resulting \texttt{memo} program directly mirrors the model
described previously.
Inside \texttt{thinks}, the buyer simulates its model of the world,
with the quality of the book being drawn from its prior and the seller
observing the quality and choosing a price according to a recursive
model \mintinline{python}{S(level, noise)}.
The buyer conditions on the price it actually observes and infers the
probability that the quality is high.
This \texttt{memo} program corresponds one-to-one to the original
\langname program, \mintinline{python}{world.choose} becomes a prior,
\mintinline{python}{seller.choose} a strategic decision, and
\mintinline{python}{send} operations delivering the observation to the
buyer.

\begin{wrapfigure}[12]{r}{0.4\columnwidth}
\vspace{-1em}
\begin{minted}[fontsize=\scriptsize]{python}
@memo
def B[q: Quality, p: Prices](
        level, noise):
    buyer: thinks[
        world: given(q in Quality,
            wpp=0.6 if q>100 else 0.4),
        seller: knows(world.q),
        seller: chooses(p in Prices,
            wpp=S[world.q, p](
                {level}, {noise}))
    ]
    buyer: observes[seller.p] is p
    buyer: chooses(q in Quality,
        wpp=Pr[world.q == q])
    return Pr[buyer.q == 1]
\end{minted}
\vspace{-1em}
\caption{Buyer Theory of Mind}
\label{fig:memo-buyer}
\end{wrapfigure}
Once the model is constructed, we can then run inference over it to
obtain a decision policy that can be used to run the protocol --- a
mapping from observed prices to a probability of accepting.
When we run inference over this protocol, we are able to recreate
aspects of the phenomenon of the marketplace for lemons~\cite{Akerlof70}:
with a depth 0 reasoning, the buyer naively assumes that high prices
strongly signal high quality goods, and accepts proportional to the
price given; the seller recursively realises this and picks high
prices irrespective of the quality of the good. As the depth of the
recursive reasoning is increased, the acceptance probabilities even
out as the buyer becomes more strategic, learning that high prices
signal quality, but with diminishing confidence.



\section{The Landscape of Multi-Agentic Ecosystems}
\label{sec:related}
In this work, we propose extending choreographic languages with
operations to encode participants' preferences and utilities, allowing
choreographies to encode not only an executable specification of
\emph{what} agents will do, but also \emph{why} participants would
want to participate.
In this way, our formalism bridges from programming languages, to a
rich literature from the field of game theory.

A number of prior works have investigated the design of DSLs to
describe specific classes of formal games.
Several DSLs target specification of auction-style games where a
seller collects bids from several participants:
MIND~\cite{yangMINDMarketInterpretation2025} provides a typed
declarative grammar for market design and simulation,
CoorERE~\cite{hoseindoostExecutableDomainspecificModeling2024} is a
graphical modelling language for auctions and De~Jonge and
Zhang~\cite{dejongeGDLUnifyingDomain2021} show that the Game
Description Language~\cite{loveGeneralGamePlaying} can encode a broad
class of auction-style negotiation domains.
Beyond auctions, Slice~\cite{bertramCuttingCakeLanguage2023} provides
a DSL for describing cake-cutting protocols.
These works have shown that language design can be used to
enforce and guide the designs of games, but each addresses a narrow
kind of game, eliding analyses of communication structure or
cryptographic tools for protocol enforcement.
No prior work provides a unified formal object that autonomous agents
can use to reason about an interaction end-to-end.

There is a parallel line of work from the multi-agent systems
literature that has explored negotiation extensively, but largely in
unstructured settings~\cite{MultiagentAISystems2026}.
In contemporary LLM-based multi-agent
frameworks~\cite{du2023improving, chanChatEvalBetterLLMbased2023,
  hongMetaGPTMetaProgramming2024}, agents coordinate through natural
language conversation: they debate, persuade, and reach consensus on
their interactions through iterative dialogue with each other.
This flexibility is appealing but fundamentally insufficient for
robust coordination.
Natural language agreements are ambiguous, unenforceable, and
vulnerable to the very failure modes outlined earlier: sycophancy,
prompt injection, and coercion.
For trustworthy coordination, the output of negotiation must be a
formal object, a protocol specification that can be analysed,
verified, and executed with guarantees, both at the substrate-level
but also \emph{security policies}.
These claims are supported by existing work on protecting single-agent
LLMs from prompt injection attacks~\cite{debenedetti2025defeating},
where prompts are turned into programs to avoid contamination from
untrusted inputs.


\section{Conclusion \& Future Work}
\label{sec:conclusion}
In this work we have presented \langname, a choreographic language
tailored for the domain of multi-agentic coordination, extending
choreographies so that agents can not only express their communication
patterns, the flow of information through a protocol, but also their
decisions, priors and utilities.
By making strategic concerns explicit, every \langname protocol
corresponds to a formal game, and game-theoretic analyses, such as the
theory of mind solver from Section~\ref{sec:memo}, can be implemented
as endpoint projections.
We have implemented \langname as an embedded DSL within Python on top
of the \texttt{effectful}~\cite{effectful2024} library, using
algebraic effects~\cite{shen2024verifiedcp} to implement endpoint
projection.
This work is preliminary, and our next steps are to formalise the
semantics of \langname, and prove the correctness of endpoint
projection to game-theoretic models.
We are also investigating richer analyses that can be built on top of
this language, such as incentive compatibility checking, equilibrium
computation and automatic mechanism design over \langname protocols by
program synthesis.


\bibliographystyle{ACM-Reference-Format}
\bibliography{references}

\end{document}